\documentclass[conference]{IEEEtran}
\IEEEoverridecommandlockouts
\usepackage{amssymb,amsmath}
\usepackage{lastpage}
\usepackage{graphicx}
\usepackage{color}
\usepackage{inputenc}
\usepackage{algorithm}
\usepackage{bigints}
\usepackage{algpseudocode}
\usepackage{multirow}
\usepackage{comment}
\usepackage{booktabs}
\usepackage[font=small, labelfont=bf]{caption}
\usepackage{url}
\usepackage{cleveref}
\usepackage{cite}
\usepackage{textcomp}
\usepackage{bold-extra}
\usepackage[]{todonotes}
\begin{document}
\title{\huge{{\sc Delmu}: A Deep Learning Approach to Maximising the Utility of Virtualised Millimetre-Wave Backhauls}}



\author{Rui Li, Chaoyun Zhang, Paul Patras, Pan Cao, and John S. Thompson\thanks{R. Li, C. Zhang, and P. Patras are with School of Informatics, University of Edinburgh, UK. Email: \{rui.li, chaoyun.zhang, paul.patras\}@ed.ac.uk. }\thanks{P. Cao is with School of Engineering and Technology, University of Hertfordshire, UK. Email: p.cao@herts.ac.uk}\thanks{J. S. Thompson is with School of Engineering, University of Edinburgh, UK. Email: John.Thompson@ed.ac.uk}}

\maketitle

\begin{abstract}
Advances in network programmability enable operators to `slice' the physical infrastructure into independent logical networks. By this approach, each network slice aims to accommodate the demands of increasingly diverse services. 
However, precise allocation of resources to slices across future 5G millimetre-wave backhaul networks, to optimise the total network utility, is challenging. This is because the performance of different services often depends on conflicting requirements, including bandwidth, sensitivity to delay, or the monetary value of the traffic incurred. 
In this paper, we put forward a general rate utility framework for slicing mm-wave backhaul links, encompassing all known types of service utilities, i.e. logarithmic, sigmoid, polynomial, and linear. We then introduce {\sc Delmu}, a deep learning solution that tackles the complexity of optimising non-convex objective functions built upon arbitrary combinations of such utilities. Specifically, by employing a stack of convolutional blocks, {\sc Delmu} can learn correlations between traffic demands and achievable optimal rate assignments. We further regulate the inferences made by the neural network through a simple `sanity check' routine, which guarantees both flow rate admissibility within the network's capacity region and minimum service levels. The proposed method can be trained within minutes, following which it computes rate allocations that match those obtained with state-of-the-art global optimisation algorithms, yet orders of magnitude faster. This confirms the applicability of {\sc Delmu} to highly dynamic traffic regimes and we demonstrate up to 62\% network utility gains over a baseline greedy approach.

\end{abstract}

\section{Introduction}

The 5\textsuperscript{th} generation mobile networks (5G) embrace~a~new wave of applications with distinct performance requirements~\cite{alliance20155g}. For example, ultra-high definition video streaming and immersive applications (AR/VR) typically demand very high data throughput. Autonomous vehicles and remote medical care are stringently delay-sensitive, belonging to a new class of Ultra-Reliable Low-Latency Communications (URLCC) services~\cite{schulz-commag-latencyiot}. In contrast, Internet of Things (IoT) applications, including smart metering and precision agriculture, can be satisfied with a best-effort service. In order to simultaneously meet such diverse performance requirements, while enabling new verticals, mobile network architectures are adopting a~\emph{virtually sliced} paradigm~\cite{3gpp23501}. 
The core idea of slicing is to partition physical network infrastructure into a number of logically~isolated networks, i.e. slices. Each slice corresponds to a specific service type, which may potentially belong to a certain tenant operator.

At the same time, cellular and Wi-Fi base stations (BSs) are deployed massively, in order to increase network capacity and signal coverage. Millimetre wave (mm-wave) technology is becoming a tangible backhauling solution to connect these BSs to the Internet in a wireless fashion at multi-Gbps~speeds~\cite{dehos2014millimeter}. 
In particular, advances in narrow beam-forming and multiple-input multiple-output (MIMO) communications mitigate the severe signal attenuation characteristic to mm-wave frequencies and respectively multiply achievable link capacities~\cite{hur2013millimeter}.


\begin{figure}[!t]
	\begin{center}
 	\includegraphics[width=\columnwidth]{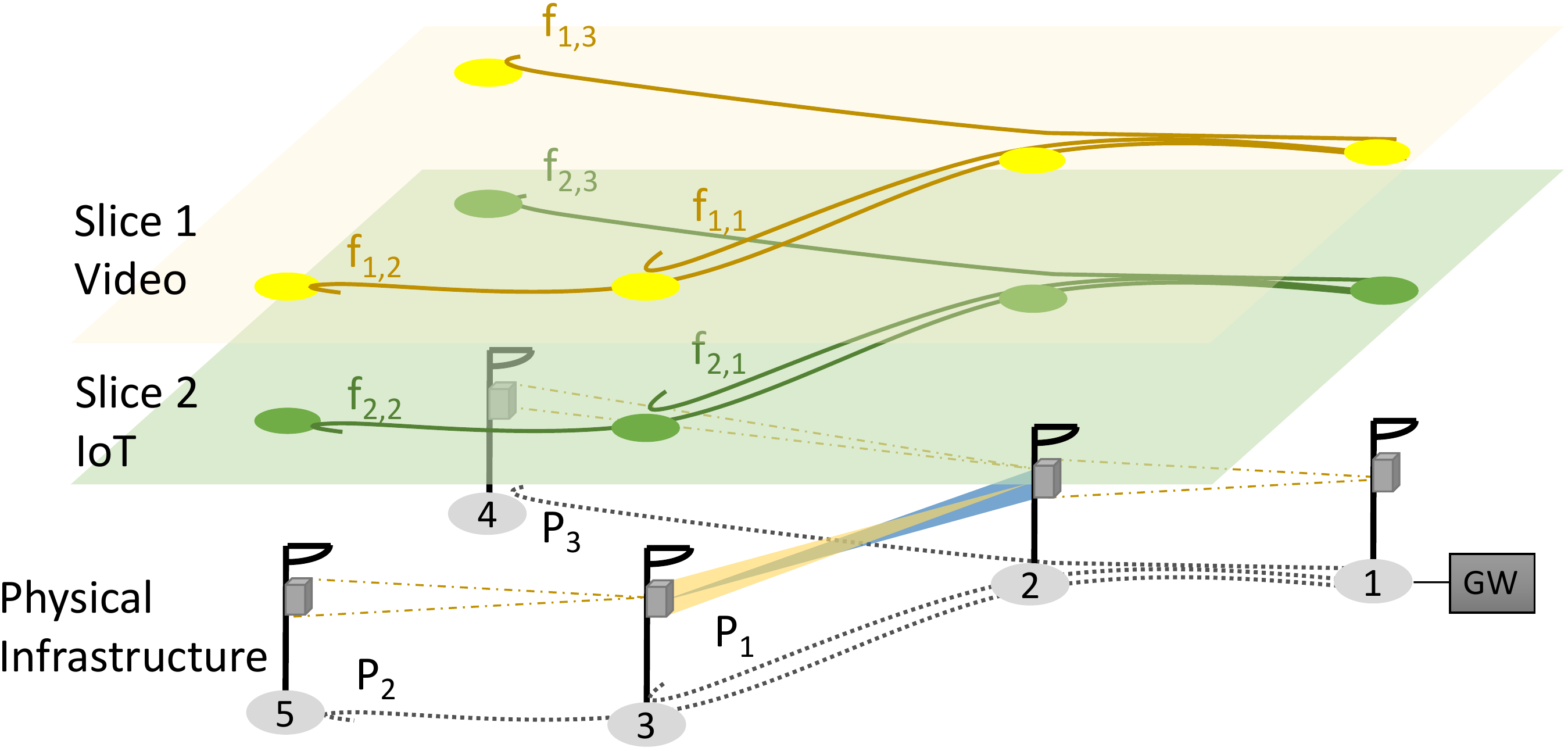}
 	\caption{Example of sliced backhaul over physical lamppost based mm-wave infrastructure. Slice 1 accommodates video streaming flows with sigmoid utilities and Slice 2 carries traffic from IoT applications, which have logarithmic utility.}
 	\label{fig:egtopo}
 	\vspace*{-2em}
	\end{center}
\end{figure}

Partitioning sliced mm-wave backhauls, and in general backhauls that employ any other communications technology, among traffic with different requirements, as in the example shown in Fig.~\ref{fig:egtopo}, is essential for mobile infrastructure providers (MIPs). By and large, MIPs aim to extract as much value as possible from network resources, yet achieving this in sliced backhauls is not straightforward. In this example, five BSs are inter-connected via mm-wave directional links, forming a shared backhaul. 
The notion of rate utility is widely used to quantify the worth of an allocation of resources to multiple flows. The question is: \emph{what type of utility is suitable to such multi-service scenarios?} Logarithmic utility as proposed in~\cite{Kelly:1997} has been adopted for elastic services and remains suitable for best-effort IoT traffic. On the other hand, applications such as video streaming typically throttle below a threshold, whilst an increase in service level is mostly imperceptible by users when the allocated rate grows beyond that threshold. Hence, the utility of such traffic can be modelled as a step-like sigmoid~\cite{Yin2015}. Had there been real-time applications to accommodate, their utility is typically formulated through polynomial functions~\cite{Fazel2005,Wang2017}. Further, in the case of traffic for which the MIP allocates resources solely based on monetary considerations, a linear utility function can be employed. However, as the application scenarios diversify, a single type of utility cannot capture the distinct features of different service types. Therefore, we argue that a mixed utility must be considered. 
Unfortunately, combining all these utility functions may lead to non-concave expressions and computing in a timely manner the optimal rate allocation that maximises their value becomes a challenging task. Global search metaheuristics explore the feasible solution space intelligently to find global maxima~\cite{ugray2007scatter}, yet often involve unacceptably long computational times. Thus they fail to meet 5G specific delay requirements in highly dynamic environments, where application demands change frequently. Greedy approaches can be used to overcome the runtime burden, though these will likely settle on sub-optimal solutions.

\textbf{Contributions:}
In this paper, we first put forward a general utility framework for sliced backhaul networks, which incorporates all known utility functions. We show that finding solutions to the network utility maximisation (NUM) problem when arbitrarily combining different utility functions is NP-hard. Inspired by recent advances in deep learning, we tackle complexity by proposing {\sc Delmu}, a deep neural network model that learns the relations between traffic demands and optimal flow rate allocations. Augmented with a simple post-processing algorithm that ensures minimum service levels and admissibility within the network's capacity, we show that {\sc Delmu} makes close-to-optimal inferences while consuming substantially shorter time as compared to state-of-the-art global search and a baseline greedy algorithm. In view of the current technological trends, we particularly focus on backhauls that operate in mm-wave bands. However, our utility framework and deep learning approach can be applied to other systems that operate in microwave or sub-gigahertz bands.

The remainder of the paper is structured as follows. In Sec.~\ref{sec:sysmodel} we discuss the system model and in Sec.~\ref{sec:problemformulation} we formulate the general NUM problem in the context of sliced mm-wave backhauls. We present the proposed deep learning approach to solving NUM in Sec.~\ref{sec:method} and show its performance in Sec.~\ref{sec:results}. We review relevant related work in Sec.~\ref{sec:relatedwork}. Finally, in Sec.~\ref{sec:conclusions} we conclude the paper.

\section{System Model}
\label{sec:sysmodel}
We consider a backhaul network deployment with $\cal{B}$ base stations (BSs) inter-connected via mm-wave links.\footnote{Although we primarily focus on mm-wave backhauls, due to their potential to support high-speed and low latency communications, the optimisation framework and deep learning solution we present next are generally applicable to other technology.} Each BS is equipped with a pair of transceivers, hence is able to transmit and receive simultaneously, while keeping the footprint small to suit dense deployment. To meet carrier-grade requirements and ensure precise TX/RX beam coordination, the network operates with a time division multiple access (TDMA) scheme. We assume carefully planned deployments where BSs have a certain elevation, e.g. on lampposts, hence interference is minimal and blockage events occur rarely.

We focus on settings where the backhaul network is managed by a single MIP and is partitioned into $I$ logical slices 
to decouple different services (e.g. as specified in~\cite{3gpp23501}). $\cal{F}$ user flows traverse the network and are grouped by traffic type~$i$ corresponding to a specific slice, i.e. ${\cal F} = \cup_{i \in \{1, ...,I\}} {\cal F}_i$.
The MIP's goal is to adjust the flow rates according to corresponding demands, in order to maximise the overall utility of the backhaul network. Flow demands are defined by upper and lower bounds. Lower bounds guarantee minimum flow rates, so as to ensure service availability, whilst upper bounds eliminate network resources wastage. We assume a controller (e.g. `network slice broker' \cite{samdanis:2016}) has complete network knowledge, periodically collects measurements of flow demands from BSs, solves NUM instances, and distributes the flow rate configurations corresponding to the solutions obtained. 

\textbf{Link Capacity:}
To combat the severe path loss experienced at mm-wave frequencies and boost capacity, BSs employ multiple input multiple output (MIMO) antenna arrays.
We consider $K$ array elements deployed at each base station for TX/RX. In backhaul settings, the stations' locations are fixed and the channel coherence time is typically long; hence it is reasonable to assume full knowledge of the channel state information is available at both transmitter and receiver sides. Given the channel matrix $\mathbf{H_{m,n}}$ from BS $m$ to BS $n$, the received signal at BS $n$ can be computed as 
\vspace*{-0.3em}
\begin{align}
\mathbf{y_n} = \mathbf{H_{m,n}x_m} + \mathbf{n_{m,n}}, 
\vspace*{-0.3em}
\end{align}
where $\mathbf{x_m}$ is an $K$-dimensional signal transmitted by BS $m$, and $\mathbf{y_n}$ are the received symbols at BS $n$. 
The singular value decomposition (SVD) of $\mathbf{H_{m,n}}$ is:
\vspace*{-0.2em}
\begin{align}
\mathbf{H_{m,n}} = \mathbf{U_n\Sigma V_m^H},
\vspace*{-0.3em}
\end{align}
where $\mathbf{U_n}$ and $\mathbf{V_m}$ are $K \times K$ unitary matrices, i.e. $\mathbf{U_nU_n^H} = \mathbf{I}$ and $\mathbf{V_mV_m^H} = \mathbf{I}$, and $\mathbf{\Sigma}$ is an $K \times K$ non-negative diagonal matrix containing the singular values of $\mathbf{H_{m,n}}$. The $k$-th diagonal entries of $\mathbf{\Sigma}$, i.e. $\sigma_k$, represents the $k$-th channel gain, and is also the $k$-th non-negative square root of the eigenvalues of matrix $\mathbf{H_{m,n}H_{m,n}^H}$.

The parallel channel decomposition can be implemented efficiently for mm-wave systems as follows~\cite{ayach:2014}. The transmitter precoding performs a linear transformation on the input vector $\mathbf{\tilde{x}_m}$, i.e. $\mathbf{x_m} = \mathbf{V_m\tilde{x}_m}$, and the received signal $\mathbf{y_n}$ is linearly decoded by $\mathbf{U_n^H}$, i.e. $\mathbf{\tilde{y}_n} = \mathbf{U_n^H y_n}$.  
Therefore, the link capacity $c_{m,n}$ between base station $m$ and $n$ can be computed as:
\begin{align}
&c_{m,n} := \max_{\substack{\mathbf{Q_m}:\\Tr(\mathbf{Q_m}) \leq P_{\max}}} B\log_2 \det(\mathbf I + \mathbf{H_{m,n}Q_mH_{m,n}^H}) \label{eq:capaH},
\end{align}
where $\mathbf{Q_m} = \mathbf{V_mV_m^H}$ is the transmission covariance matrix, $B$ is the channel bandwidth, and $P_{\max}$ is the maximum transmit power. Without loss of generality, we assume that all BSs have the same maximum transmit power budget.

For a channel known at the transmitter, the optimal capacity can be achieved by the well-known channel diagonalisation and the water-filling power allocation method~\cite{raleigh:1998}. For all BS $m$, by employing the optimal transmit pre-coding matrix $\mathbf{V_m = X_{m,n}}$, where $\mathbf{X_{m,n}}$ denotes the eigenvector matrix of $\mathbf{H_{m,n}^H*H_{m,n}}$, the MIMO channel capacity maximisation can be reformulated as:
\begin{align}
&c_{m,n} := \max B\sum_{k=1}^K\left(\log\left(1+ \frac{\lambda_k p_m^k}{\epsilon^2}\right)\right),\\
\text{s.t.} \ &0 \leq \sum_{k=1}^K p_m^k \leq P_{\max}, \quad \label{eq:powerCons}\\
&p_m^k \geq 0 , \forall k, m\quad
\end{align}
where $\lambda_k = \sigma_k^2$, and $\epsilon^2$ denotes the noise power. If the power allocated on the~\mbox{$k$-th} sub-channel is $p_m^k$ at BS $m$, then (\ref{eq:powerCons}) specifies the total transmit power constraint. The optimal water-filling power allocation yields $p_m^k = \max \{0, \mu - \epsilon^2/\lambda_k\}$, where $\mu>0$ is the water-filling level such that $\sum_{k=1}^K p_m^k = P_{\max}, \forall m$~\cite{raleigh:1998}.


\section{Problem Formulation}
\label{sec:problemformulation}
Our objective is to find the optimal end-to-end flow rates that maximise the utility of sliced multi-service mm-wave backhaul networks. We first introduce a general network utility framework, based on which we formulate the NUM problem, showing that in general settings this is \mbox{NP-hard}. 

\subsection{Utility Framework}
Recall that network utility refers to the value obtained from exploiting the network, which can be monetary, resource utilisation, or level of user satisfaction. For any flow $f$ we consider four possible types of utility functions of flow rate $r$, depending on which slice ${\cal F}_i$ that flow belongs to. The utilities considered are parameterised by $\alpha_i$ and $\beta_i$, whose values have practical implications, such as the amount billed by the MIP for a service. Given an allocated rate $r$, we distinguish the following types of services that can be mapped onto slices, whose utilities we incorporate in our framework:
\begin{enumerate}
\setlength\itemsep{0.8em}
\item Services for which the MIP aims to maximise solely the attainable \textbf{revenue}. Denoting ${\cal F}_1$ the set of flows in this class, their utility is formulated as a linear function~\cite{Ahuja:1993}:
\begin{align}
U_{\text{lnr}}(r) = \alpha_{1} r + \beta_1, \quad \forall{f \in {\cal F}_1}.
\end{align}
We note that $U_{\text{lnr}}(r)$ is both concave and convex.
\item Flows $f \in {\cal F}_2$ generated by applications that require certain level of \textbf{quality of service}, e.g. video streaming, and whose corresponding utility is thus formulated as a sigmoid function~\cite{Yin2015}:
\begin{align}
U_{\text{sig}}(r) = \cfrac{1}{1+ e^{- \alpha_{2} (r - \beta_2)}}, \quad \forall{f \in {\cal F}_2}.
\end{align}
Observe that $U_{\text{sig}}(r)$ is convex in $[0,\beta_2)$ and concave in $(\beta_2, \infty)$, therefore non-concave over the entire domain. 
\item \textbf{Delay sensitive} flows, $f \in {\cal F}_3$, whose utility is modelled as a polynomial function~\cite{Fazel2005}:
\begin{align}
U_{\text{ply}}(r) = \alpha_{3} (r ^{\beta_3}), \quad \forall{f \in {\cal F}_3}, 
\end{align}
where $\beta_3$ is in the range $(0,1]$, for which the above expression is concave. 

\item \textbf{Best-effort} traffic, $f \in {\cal F}_4$, that does not belong in any of the previous classes, and whose utility is commonly expressed through a logarithmic function~\cite{Kelly:1997}:
\begin{align}
U_{\text{log}}(r) = \log (\alpha_{4} r +\beta_4), \quad \forall{f \in {\cal F}_4}. 
\end{align}
It is easy to verify that $U_{\text{log}}(r)$ is also concave.
\vspace*{0.2em}
\end{enumerate}

\begin{figure}[!t]
	\begin{center}
 	\includegraphics[width=0.9\columnwidth]{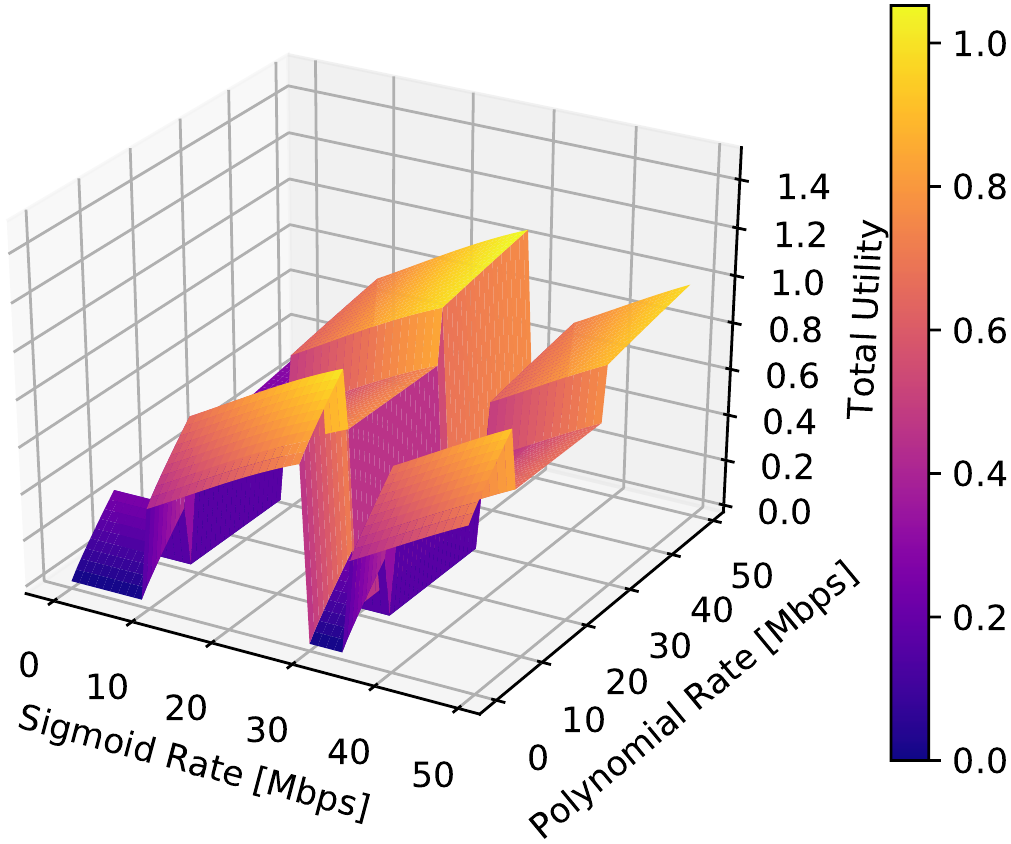}
 	\caption{Total utility when combining four flows with different utility functions; namely, two have sigmoid utility parametrised by $\alpha_2 = 0.08$, $\beta_2 = 15$ in the $[10-30]$~Mbps range, and respectively $\alpha_2 = 0.08$, $\beta_2 = 40$ in the $[35-50]$~Mbps range; the other two flows have polynomial utility with $\alpha_3 = 0.03651$, $\beta_3 = 0.9$ between $[0-10]$~Mbps, and $\alpha_3 = 0.03$, $\beta_3 = 0.6$ in $[30-50]$~Mbps. Rates increased in tandem for each type of flow. }
 	\label{fig:funcs}
	\end{center}
\vspace*{-1em}
\end{figure}

Our general utility framework encompasses all the four types of traffic discussed above (which may be parametrised differently for distinct tenants), therefore we express the overall utility of the sliced backhaul network as
\begin{align}
 & {\cal U} := \sum_{f\in {\cal F}} U(r) = \sum_{f_1\in {\cal F}_1} U_{\text{lnr}}(r_1) + \sum_{f_2\in {\cal F}_2} U_{\text{sig}}(r_2) \nonumber\\
 &\quad + \sum_{f_3\in {\cal F}_3} U_{\text{ply}}(r_3) + \sum_{f_4\in {\cal F}_4} U_{\text{log}}(r_4) \label{totalU}.
\end{align}
Arbitrary combinations of both concave and non-concave utility functions may result in non-concave expressions ${\cal U}$, as exemplified in Fig.~\ref{fig:funcs}. In this figure, we show the total utility when combining 4 flows with different utility functions, two of them sigmoidal and two polynomial, each with different parameters. We assume the rates of each type of flow increase in tandem. Observe that even in a simple setting like this one, the network utility is highly non-concave and finding the optimal allocation that maximises it is non-trivial. We next formalise this problem with practical mm-wave capacity constraints, following which we discuss its complexity.

\subsection{Network Utility Maximisation}
Consider a set of flows that follow predefined paths, $P_j, j \in\{1,2, ... ,J\}$, to/from the local gateway, where the number of possible routes in the network is $J$. We denote $f_{i,j}$ a flow on slice $i$ that traverses path $P_j$, which is allocated a rate $r_{i,j}$. By contract, $r_{i,j}$ shall satisfy $\delta_{i,j} \leq r_{i,j} \leq d_{i,j}$, where $\delta_{i,j}$ is the minimum rate that guarantees service availability, and $d_{i,j}$ is the upper bound beyond which the service quality cannot be improved. $d_{i,j}$ is no less than $\delta_{i,j}$ by default.
Furthermore, each path $P_{j}$ consists of a number of mm-wave links, and the link between BSs $m$ and $n$ is subject to a link capacity $c_{m,n}$. We use $\tau_{j,m}^{\text{s}} \in \{0,1\}, s \in \{Tx, Rx\}$, to indicate whether node $m$ transmits or receives data of flows traversing path $P_{j}$. 
The total network utility in (\ref{totalU}) can be rewritten as:
\vspace*{-0.5em}
\begin{align}
 & \sum_{f\in {\cal F}} U(r) = \sum_{i=0}^I\sum_{j=0}^J U_i(r_{i,j}). 
 \vspace*{-0.5em}
\end{align}

Finding the flow rate allocation vector $\mathbf{r_{i,j}}, \forall i, j$, that maximises this utility requires to periodically solve the following optimisation problem:
\begin{align}
 & \max \sum_{i=0}^I\sum_{j=0}^J U_i(r_{i,j})  \label{func:object}\\
 \hspace*{-0.5em} \text{s.t.}\ & \delta_{i,j} \leq r_{i,j} \leq d_{i,j}, \forall i,j; \label{constraint1}   \\
	         & \sum_{i=0}^I\sum_{j=0}^J \tau_{j,m}^{s}\frac{r_{i,j}}{c_{m,n}} \leq 1, \{m,n\} \in P_j, s \in \{Tx, Rx\}. \label{constraint2} 
\end{align}
In the formulation above, (\ref{func:object}) is the overall objective function and (\ref{constraint1}) specifies the demand constraints.
Each BS can transmit and receive to/from one and only one BS simultaneously, and the total time allocated at a single node for all flow Tx/Rx should not exceed 1, which is captured by (\ref{constraint2}).
Here $r_{i,j}/{c_{m,n}}$ denotes the time fraction allocated to flow $f_{i,j}$ on link $l_{m,n}$. 

\subsection{Complexity}
In what follows we briefly show that the network utility optimisation problem formulated above, where the objective function is a linear combination of linear, sigmoid, polynomial, and logarithmic functions, is NP-hard. By Udell and Boyd~\cite{Udell2013} any continuous function can be approximated arbitrarily well by a suitably large linear combination of sigmoidal functions~\cite{Udell2013}. Thus $\sum U(r)$ can be regarded as a sum of sigmoids and a larger number of other sigmoidal functions. Following the approach in~\cite{Udell2013}, we can reduce an integer program 
\begin{align*}
 &\text{find} \ \mathbf{r} \\
    \text{s.t.}\ & A\mathbf{r} \leq Z;~\mathbf{r} \in \{0,1\} ^n,
\end{align*}

\noindent to an instance of a sigmoidal program
\begin{align*}
  & \max \sum_{i} g(r_i) = \sum_{i} r_i(r_i-1)\\
    \text{s.t.}\ & A\mathbf{r} \leq Z;~0 \leq r_i \leq 1.
\end{align*}
Here $g(r_i)$ enforces a penalty on non-integral solutions, i.e. the solution to the sigmoidal program is 0 if and only if there exists an integral
solution to $A\mathbf{r} = Z$. Since the integer program above is known to be NP-hard~\cite{papadimitriou1998combinatorial}, the reduced sigmoid program is also NP-hard, and therefore the NUM problem we cast in (\ref{func:object})--(\ref{constraint2}) is also NP-hard.

\section{{\sc Delmu:} A Deep Learning Approach to NUM}
\label{sec:method}
To tackle the complexity of the optimisation problem formulated in the previous section and compute solutions in a timely manner, we propose {\sc Delmu}, a deep learning approach specifically designed for sliced mm-wave backhauls and also applicable to other technologies. In essence, our proposal learns correlations between traffic demands and allocated flow rates, to make inferences about optimal rate assignments. We show that, with sufficient training data, our deep neural network finds solutions close to those obtained by global search, while requiring substantially less runtime. 

\subsection{Convolutional Neural Network}
\begin{figure}[t]
	\begin{center}
 	\includegraphics[width=\columnwidth]{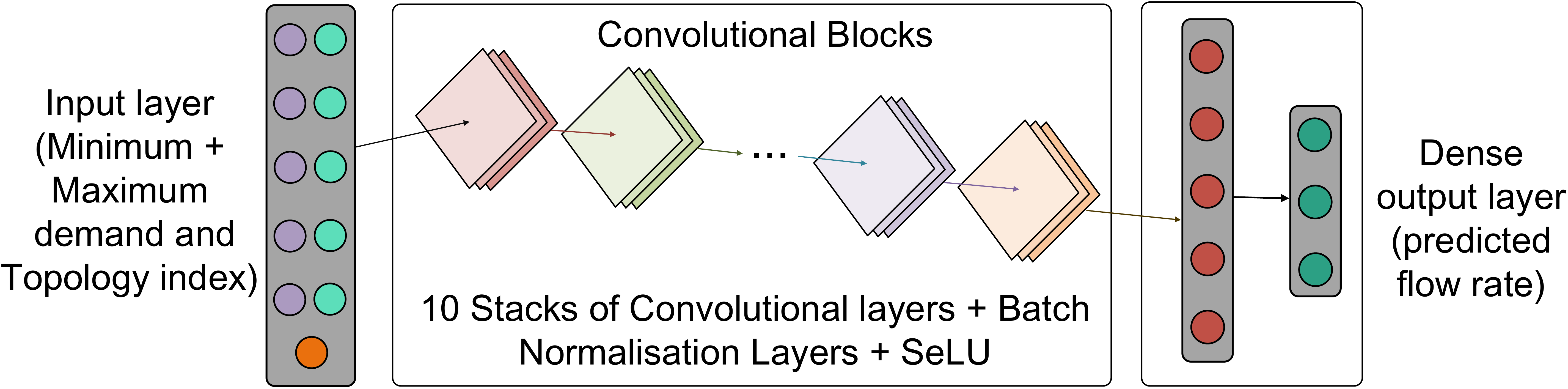}
 	\caption{Proposed Convolutional Neural Network with 10 hidden layers, which takes traffic demand and topology index as input, and infers the optimal flow rate~allocations.}
 	\label{fig:nn}
 	\vspace*{-1em}
	\end{center}
\end{figure}

We propose to use a Convolutional Neural Network (CNN) to imitate the behaviour of global search. We train the CNN by minimising the difference between ground-truth flow rates allocations (obtained with global search) and those inferred by the neural network. In general CNNs preform weight sharing across different feature channels~\cite{goodfellow2016deep}. This significantly reduces the number of model parameters as compared to traditional neural networks, while preserving remarkable performance. At the same time, our approach aims to work well with a limited amount of training data, which makes CNNs particularly suitable for our problem. Therefore, we design a 12-layer CNN to infer the optimal flow rate and illustrate its structure in Fig.~\ref{fig:nn}. The choice is motivated by recent results that confirm neural network architectures with 10 hidden layers, like ours, can be trained relatively fast and perform excellent hierarchical feature extraction~\cite{srivastava2015training}.

The minimum and maximum traffic demand, and topology information are concatenated into a single vector, which will be subsequently fed to a sequence of convolutional blocks. Each block consists of a one-dimensional convolutional layer and a Scaled Exponential Linear Unit (SELU)~\cite{klambauer2017self}, which takes the following form:
\begin{equation}
\mathrm{SELU}(x) = \omega \begin{cases}
x & x >0\\
\eta e^x - \eta & x\leqslant 0. 
 \end{cases}
\end{equation}
Here $\omega = 1.0507$ and $\eta =1.6733$ by default. Employing SELU functions aims at improving the model representability, while enabling self-normalisation without requiring external techniques (e.g. batch normalisation). This enhances the robustness of the model and eventually yields faster convergence. Features of traffic demands are hierarchically extracted by convolutional blocks, and they are sent to fully-connected layers for inference. We train the CNN using a stochastic gradient descent (SGD) based method named Adam \cite{kingma2015adam}, by minimising the following mean square error:
\begin{equation}
\textsl{L}_e = \frac{1}{Q\times I\times J}\sum _{q=0}^Q \sum _{i=0}^I \sum_{j=0}^J (r_{q,i,j} - r'_{q,i,j})^2.
\label{eq:mse}
\end{equation}
$Q$ denotes the number of training data points, $r_{q,i,j}$ denotes the allocated rate allocated to flow $j$
on slice $i$, with demand instance $q$, as suggested by global search. $r'_{q,i,j}$ is the corresponding rate inferred by the neural network. We train the CNN with 500 epochs, with an initial learning rate of 0.0001.

\subsection{Post-Processing Algorithm}
The output of the CNN on its own occasionally violates the constraints~(\ref{constraint1}) and~(\ref{constraint2}), because the model is only fed with traffic demands without embedding of constraints. We address this issue by designing a post-processing algorithm that adjusts the CNN solutions to fall within feasible domains, while maintaining minimum utility degradation and very short computation times. The idea is to first decrease recursively with a large step-length the rate of flows that breach the constraints, then increase repeatedly with a smaller step-length the rate of flows that can achieve the largest utility gains.

\begin{algorithm}[!h]
\small
\caption{CNN Post-Processing Algorithm}
\label{alg:rounding}
\begin{algorithmic}[1]
		\State{Compute the time between each pair of nodes $t_{m,n}^s$} \label{ln:time1}		
		\State{Compute the utility of each flow $u_{i,j} = U_i(r_{i,j})$}  \label{ln:utility1}  
		\While{Any $t_{m,n}^s>1$} \label{ln:while}
				\State{Find the link $l_{m,n}$ with the maximum $t_{m,n}^s$} \label{ln:findmaxtime}
				\State{$\text{deStepLen} = \min \{10,  r_{i,j} - \delta_{i,j}\}$} \label{ln:searchMinDecrease}
				\For{Flows satisfying $\tau^s_{j,m} == 1$ or $\tau^s_{j,n} == 1$ for $l_{m,n}$}
					\State{Potential utility decrease $u'_{i,j} = U_i(r_{i,j} -\text{deStepLen})$}
				\EndFor\label{ln:endfor}
				\State{Find the $f_{i,j}$ with the minimum non-zero $\Delta u_{i,j} = u_{i,j} - u'_{i,j}$}\label{ln:minloss}
				\State{Decrease rate of $f_{i,j}$, i.e. $r_{i,j} =  r_{i,j} - \text{deStepLen}$}\label{ln:decrease}
				\State{Update $t_{m,n}^s$ and $u_{i,j}$}\label{ln:utility2}
		\EndWhile			 \label{ln:endwhile}
		
		\While{Any flow rate can be increased} \label{ln:while2}
				\State{$\text{inStepLen} = \min \{1,  d_{i,j} - r_{i,j} \}$} \label{ln:inStepLen}
				\State{Potential utility increase $u''_{i,j} = U_i(r_{i,j} + \text{inStepLen}), \forall f_{i,j}$}\label{ln:uIncrease}
				\State{Find the $f_{i,j}$ with the maximum $\Delta u_{i,j} = u''_{i,j} - u_{i,j} $} \label{ln:uIncreaseMax}
				\State{Increase rate of $f_{i,j}$, i.e. $r_{i,j} =  r_{i,j} + \text{inStepLen}$} \label{ln:uIncreaseConfirm}
				\State{Update $t_{m,n}^s$ and $u_{i,j}$} \label{ln:utility3}  
		\EndWhile			 \label{ln:endwhile2}			
\end{algorithmic}
\end{algorithm}

Algorithm~\ref{alg:rounding} shows the pseudo-code of this procedure. The routine starts by computing the total time on each link for all traversing flows, i.e. $t^s_{m,n} = \sum_{i}\sum_{j} \tau_{j,m}^{s} r_{i,j}/c_{m,n}$ (line~\ref{ln:time1}) and the utility of each individual flow based on the rate allocation returned by CNN (line~\ref{ln:utility1}). Then it searches recursively for a flow to decrease (lines~\ref{ln:while}--\ref{ln:endwhile}). At each step, Algorithm~\ref{alg:rounding} selects the link with the highest total time (line~\ref{ln:findmaxtime}) and reduces the rate of the flow traversing the link with minimum possible utility loss (lines~\ref{ln:searchMinDecrease}--\ref{ln:decrease}). Then the total link time and the flow utilities are updated (line~\ref{ln:utility2}). The process (lines~\ref{ln:findmaxtime}--\ref{ln:utility2}) is repeated until the time for all links comply with the time constraints. Next, we increase iteratively a flow that yields the maximum potential utility gain, while ensuring that all constraints are satisfied (lines \ref{ln:while2}--\ref{ln:endwhile2}). This is done by tentatively increasing each flow, with a step-length that complies with the demand constraint (line \ref{ln:inStepLen}), computing the corresponding utility increment (line \ref{ln:uIncrease}), then finding the flow with maximum possible utility increase (line \ref{ln:uIncreaseMax}), and confirming the rate increment for that flow (line \ref{ln:uIncreaseConfirm}). Before the next round of increasing the rates, Algorithm~\ref{alg:rounding} recomputes the total time on all links to verify that further rate increases are possible, and updates the utility of each flow (line \ref{ln:utility3}). 


\begin{figure}[t]
	\begin{center}
 	\includegraphics[width=1\columnwidth]{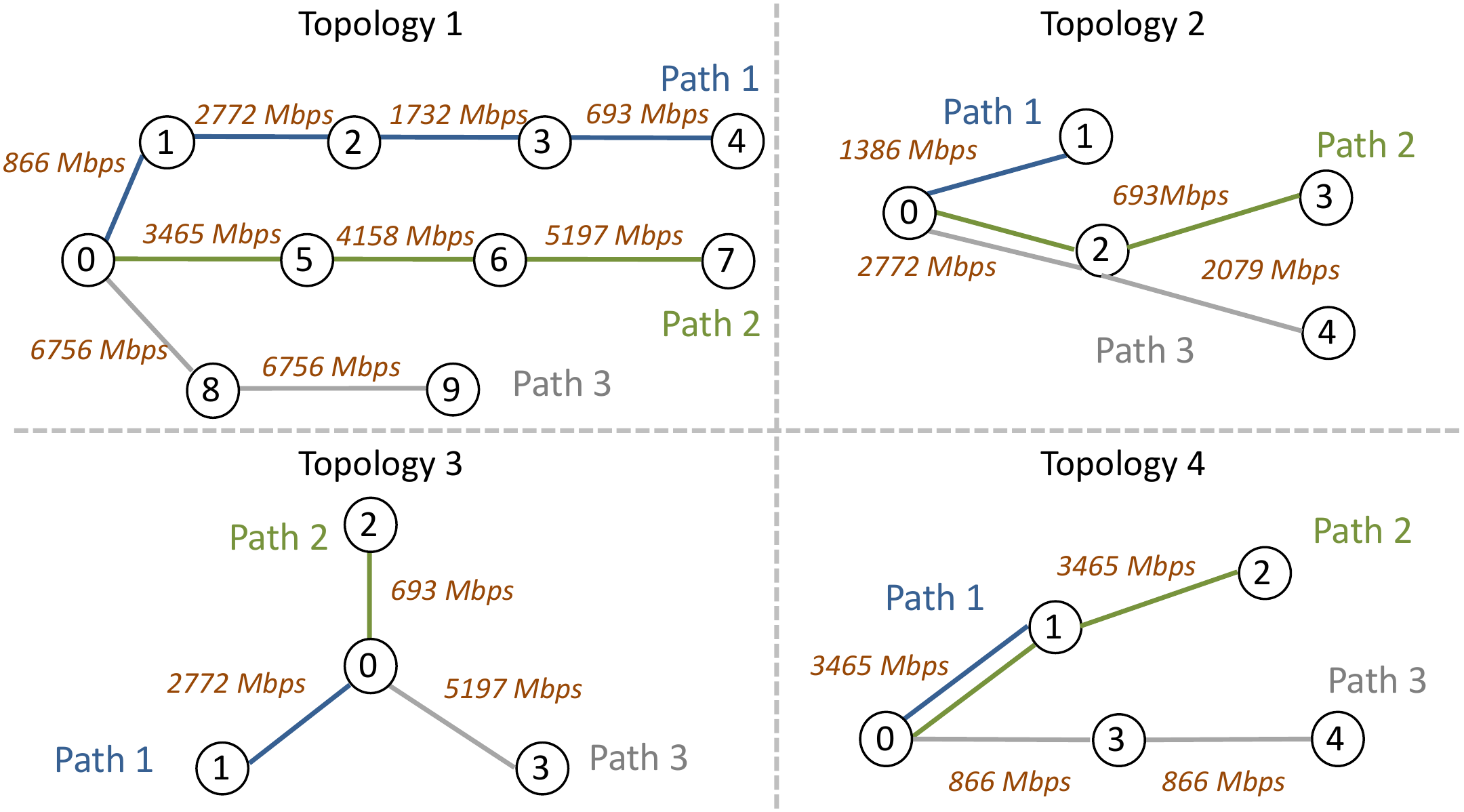}
 	\caption{The four network topologies used for evaluation. Circles represent the BSs, flow paths are shown with lines of different colour, and link capacities are labelled.}
 	\label{fig:topos}
 	\vspace*{-1.75em}
	\end{center}
\end{figure}

\section{Performance Evaluation}
\label{sec:results}
We evaluate the proposed {\sc Delmu} solution, which encompasses the CNN structure and the post processing algorithm, on different backhaul topologies under a range of conditions. Specifically, we use four different topologies as illustrated in Fig.~\ref{fig:topos}, where the number of BSs varies from 4 to 10, and link rates range from 693~Mbps to 6.8~Gbps. Each path carries up to four types of flows, i.e. with individual sigmoid, linear, polynomial, and logarithmic utilities. For each topology, we generate randomly 10,000 combinations of flow demands in the range $[0-750]$~Mbps in increments of 50~Mbps. The corresponding minimum service rates are generated uniformly at random in the range $[0-100]$ Mbps as integer values, and are capped by the maximum flow demand. The parameters shown in Table~\ref{tab:parameters} are used to model utility. 

\begin{table}[!h]
	 \centering
	 \small
	 {\begin{tabular}{c|c|c|c|c}
	  \hline
		\begin{tabular}
	     {@{}c@{}}Utility Type \end{tabular} & Linear  &Sigmoid  &Polynomial &Logarithmic\\\hline\hline	
		 $\alpha_i$  &0.00133 &0.08000 &0.03651 &0.00229  \\\hline      
		 $\beta_i$  & 0 &350 &0.5 &1  \\\hline  
	      \end{tabular}}
          
	 \caption{$\alpha_i$ and $\beta_i$ parameters for the utility functions used in the evaluation.}
	 \vspace*{-1em}
	 \label{tab:parameters}
\end{table}

To train and subsequently test the neural network, we run a global search (GS) algorithm, the optimality of which is proven in~\cite{ugray2007scatter}, on each of the 10,000 network settings described above. We use 80\% of the results obtained to construct a synthetic dataset that we use in the training process, which effectively seeks to minimise the mean square error expression defined in (\ref{eq:mse}), by means of SGD. We use the remaining 20\% of cases for as ground truth for testing the accuracy of the optimal rate allocation inferences that {\sc Delmu} makes. 
More precisely, we compare the performance of {\sc Delmu} in terms of total network utility and computational time, against the solutions obtained with GS and those computed with a baseline greedy approach that we devise. We discuss both benchmarks in more detail in the following subsection.

To compute solutions with the GS and greedy algorithms, and make inferences with the proposed CNN, we use a workstation with an Intel Xeon E3-1271 CPU @ 3.60GHz and 16GB of RAM. The CNN is trained on a NVIDIA TITAN X GPU using the open-source Python libraries TensorFlow~\cite{abadi2016tensorflow} and TensorLayer~\cite{tensorlayer}. 
We implement the greedy solution in Python and employ the GS solver of MATLAB\textsuperscript{\textregistered}.

\subsection{Benchmarks}
The GS method works by starting from multiple points within the feasible space and searching for local optima in their vicinity, then concluding on the global optimum from the set of local optima obtained~\cite{ugray2007scatter}. With default settings, which we employ in our evaluation, the GS generates 1,000 starting points using the scatter search algorithm~\cite{glover1998template}, then eliminates those starting points that are not promising (judging by the corresponding value of the objective function and constraints). It then repeatedly executes a constrained nonlinear optimisation solver, i.e.~\texttt{fmincon}, to search for local maxima around the remaining start points. Eventually the largest of all local maxima is taken as the global maximum, if one exists. We let the local optimisation routine work with the default Interior Point algorithm, which satisfies bounds at all iterations and can recover from non-numeric results. We note that simpler approximations such as semidefinite programming are constrained to convex optimisation problems, thus inappropriate for our task.

We also engineer a baseline greedy algorithm for the purpose of evaluation, with the goal of finding reasonably good solutions \emph{fast}. The greedy approach starts by setting all flow rates to the minimum demand and then recursively chooses a flow to increase its rate, with the aim of achieving maximum utility gain at the current step, as long as the constraints~(\ref{constraint1})--(\ref{constraint2}) are respected. A solution is found when there are no remaining flows whose rates can be further increased. For fair comparison, the greedy approach takes exactly the same flow demands and the corresponding minimum service rates as used by GS and {\sc Delmu}. A step size of 1 Mbps is employed.

\subsection{Total Utility}
\label{subsec:utilityResults}
\begin{figure}[t]
    \vspace*{-1em}
	\begin{center}
 	\includegraphics[width=1\columnwidth]{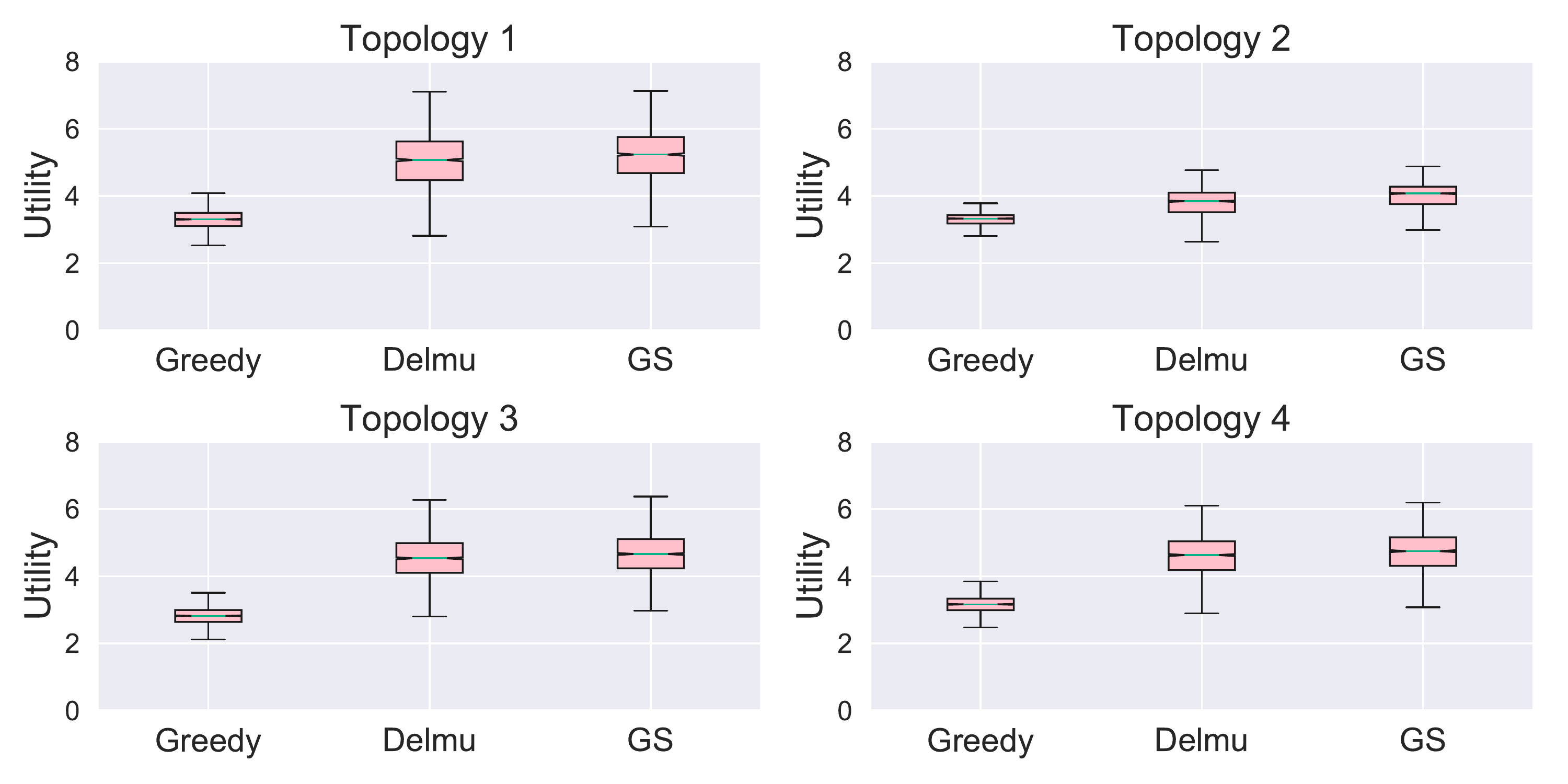}
 	\caption{Distribution of the total utility attained by the proposed {{\scshape Delmu}}, and the benchmark GS and greedy algorithms, for the four topologies shown in Fig.~\ref{fig:topos}. Numerical results.}
 	\label{fig:utilitybox}
 	\vspace*{-1em}
	\end{center}
\end{figure}

We first examine the overall utility performance of the proposed {\sc Delmu}, in comparison with that of the greedy and the GS solutions. Fig.~\ref{fig:utilitybox} illustrates the distributions of the total network utility for the 12 flows traversing the network, over the 2,000 instances tested. 
We observe that, among the 4 topologies used, the distribution of the total utility obtained by {\sc Delmu} is almost the same as that of the optimal solution obtained with GS, as confirmed by the similar median values, the distance between the first and third quartiles, as well as the whiskers (minima and maxima). 
Specifically, the median values of the total utility attained by GS in Topologies \mbox{1--4} are 5.23, 4.07, 4.66, and 4.75, while those achieved by the proposed {\sc Delmu} are 5.09, 3.88,  4.56, and 4.64. 
In sharp contrast to the {\sc Delmu}'s close-to-optimal performance, the greedy solution attains the medians of 3.30, 3.32, 2.81, and 3.16 utility units in the 4 topologies considered. Among these, for the case of Topology 3, {\sc Delmu} obtains a 62\% total utility gain over the greedy approach. It is also worth remarking that, although a greedy approach can perform within well-defined bounds from the optimum when working on submodular objective functions~\cite{SON201595}, this is clearly suboptimal in the case of general utility functions as addressed herein.

\subsection{Decomposing Performance Gains}
\begin{figure}[!t]
	\centering
 	\includegraphics[width=1\columnwidth]{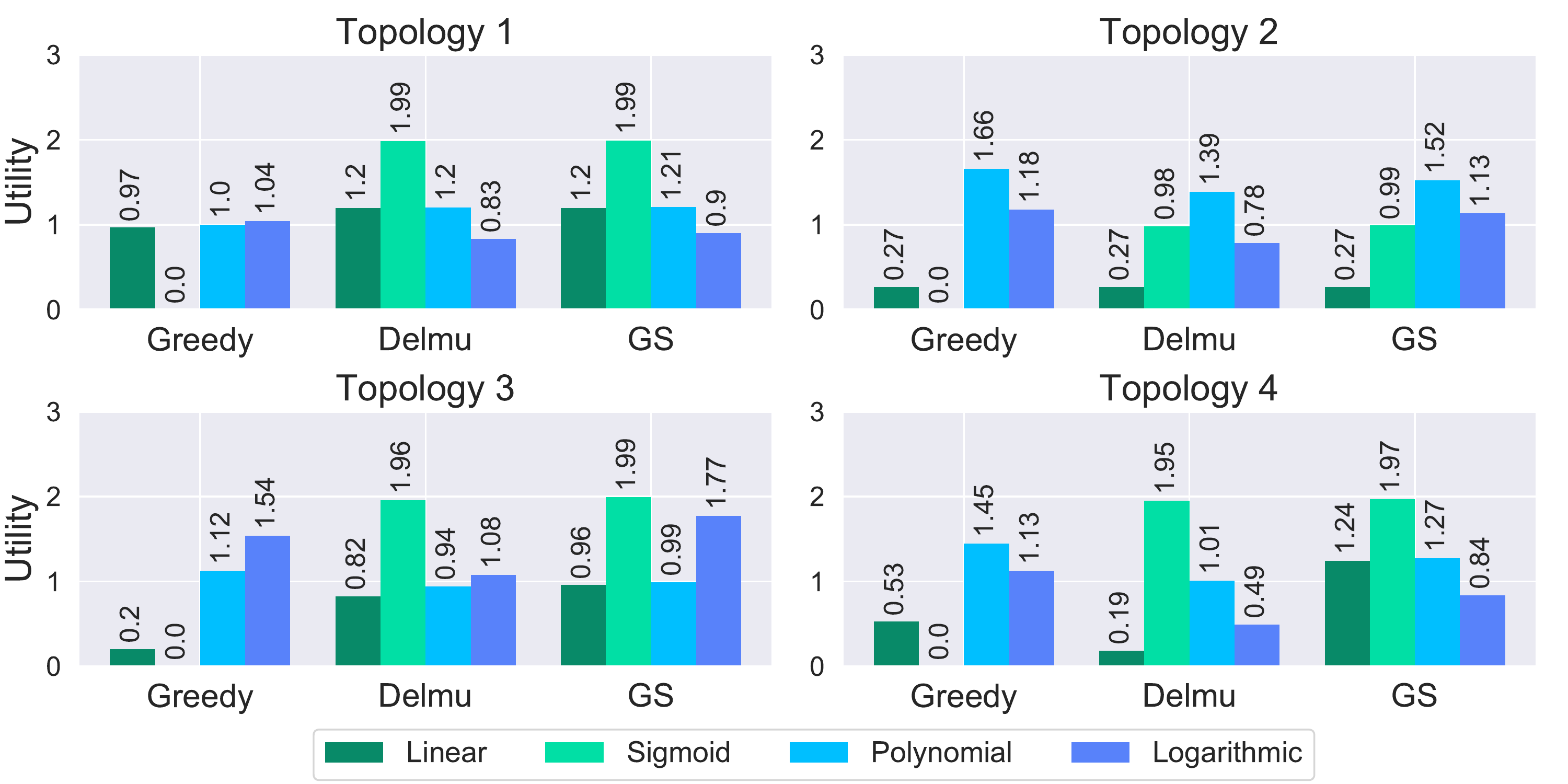}
 	\caption{An example instance of the utility corresponding to each traffic type in each topology. Bars represents the sum utility of flows in the same slice. Numerical results.}
 	\label{fig:utilitySingle}
 	\vspace*{-1em}
\end{figure}

To understand how {\sc Delmu} achieves close-to-optimal utility, and why the benchmark greedy solution performs more poorly, we examine one single instance for each topology, and dissect the utility values into the components corresponding to each type of traffic (i.e. slice). Fig.~\ref{fig:utilitySingle} illustrates the sum of utilities for each type of traffic, attained with the greedy, CNN, and GS approaches. We note that the greedy solution tends to allocate more resources to traffic with logarithmic utility (in all topologies) and respectively polynomial utility (in Topologies 2, 3, and 4). In contrast, the CNN allocates higher rates to traffic subject to sigmoid utility in all the scenarios studied, which results in higher overall utility. 
This is because the greedy approach gives more resources to the flows that yield utility gains in the first steps of the algorithm's execution and fails to capture the inflection point of the traffic with sigmoid utility, which can contribute to a higher overall utility, under limited resource constraints. 
Furthermore, the allocations of rates to different traffic types by {\sc Delmu} show close resemblance to the GS behaviour, which confirms the fact that {\sc Delmu} achieves overall close to optimal utility allocations, at a lower computational cost, as we will see next.

\begin{figure}[h]
	\centering
 	\includegraphics[width=1\columnwidth]{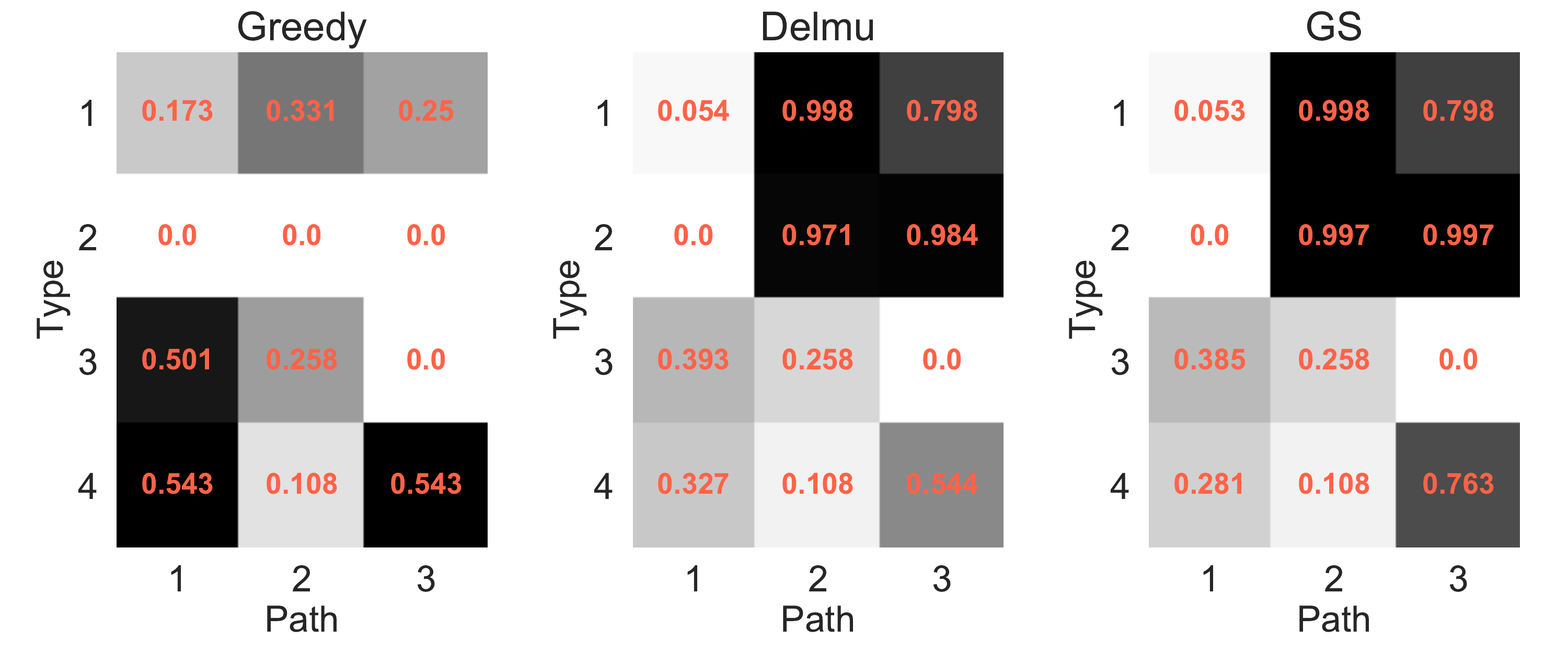}
 	\caption{Utility of all data flows (on different slices and over different paths) attained by greedy, {{\scshape Delmu}}, and GS in one demand instance in Topology 1.
 	In each subfigure, darker shades represent higher utility and the actual values are labelled. Numerical results.}
 	\label{fig:utilitySingleInstance}
\end{figure}

We delve deeper into the utility attained by each flow on each slice, along different paths, and in Fig.~\ref{fig:utilitySingleInstance} compare the performance 
of our approach and the benchmarks considered in the case of Topology 1. Flows corresponding to slices that have linear, sigmoid, polynomial, and respectively logarithmic utility are indexed from 1 to 4. Again, observe that the greedy approach assigns zero utility to traffic subject to sigmoid utility, in stark contrast with the GS method. While {\sc Delmu} obtains the highest gains from traffic with linear and sigmoid utility on paths 2 and 3, greedy dedicates most of the network resources to traffic with logarithmic and exponential utility, without obtaining significantly more utility from these types of flows. {\sc Delmu} achieves accurate inference, as the performance is nearly the same with that of GS for all flows.

\subsection{Real-time Inference}
To shed light on the runtime performance of the proposed {\sc Delmu} solution, we first examine the average time required for inferring a single solution throughout the performance analysis presented in Section~\ref{subsec:utilityResults}. We compare these computation times with those of the greedy and GS approaches over 2,000 instances and list the obtained results in Table~\ref{tab:time}.
\begin{table}[!h]
	 \centering
	  \begin{tabular}{c|l|l|l|l}
		\hline
		\begin{tabular}
		 {@{}c@{}}Topology  Index \end{tabular} & \multicolumn{1}{|c|}{1}   &\multicolumn{1}{|c|}{2}   &\multicolumn{1}{|c|}{3}    &\multicolumn{1}{c}{4} \\ \hline\hline
		GS &8.4339s   &4.6075s  &3.4492s   &4.8311s\\\hline
		Greedy &0.1500s &0.1590s &0.1178s &0.1345s\\\hline
		{\scshape {\bfseries Delmu}} &\textbf{0.0036s}   &\textbf{0.0035s}  &\textbf{0.0025s}  &\textbf{0.0026s} \\\hline     
	      \end{tabular}
     \vspace*{1em}
	 \caption{Average computation time required to obtain a single solution to the NUM problem in Topologies 1--4 using GS, greedy, and the proposed CNN mechanism.}
	 \label{tab:time}
	\vspace*{-1em}
\end{table}
Note that the values for {\sc Delmu} include the post-processing time. Observe that GS takes seconds to find a solution, while the greedy approach, although inferior in terms of utility performance, has runtimes in the order of hundreds of milliseconds for a single instance.
In contrast, our CNN makes and adjusts inferences within a few milliseconds. That is, as compared to the greedy algorithm, CNN generally requires two orders of magnitude smaller computation time. On the other hand, the GS algorithm, although working optimality, has three orders of magnitude higher runtimes as compared to {\sc Delmu}. Lastly, note that the CNN inference itself requires $\sim$1.5ms per instance, and hence the post-processing dominates the overall execution time in the first two topologies. We conclude that the proposed {\sc Delmu} is suitable for highly dynamic backhauls.

\begin{figure}[t!]
	\centering
    \vspace*{-1.5em}
	\includegraphics[width=1\columnwidth]{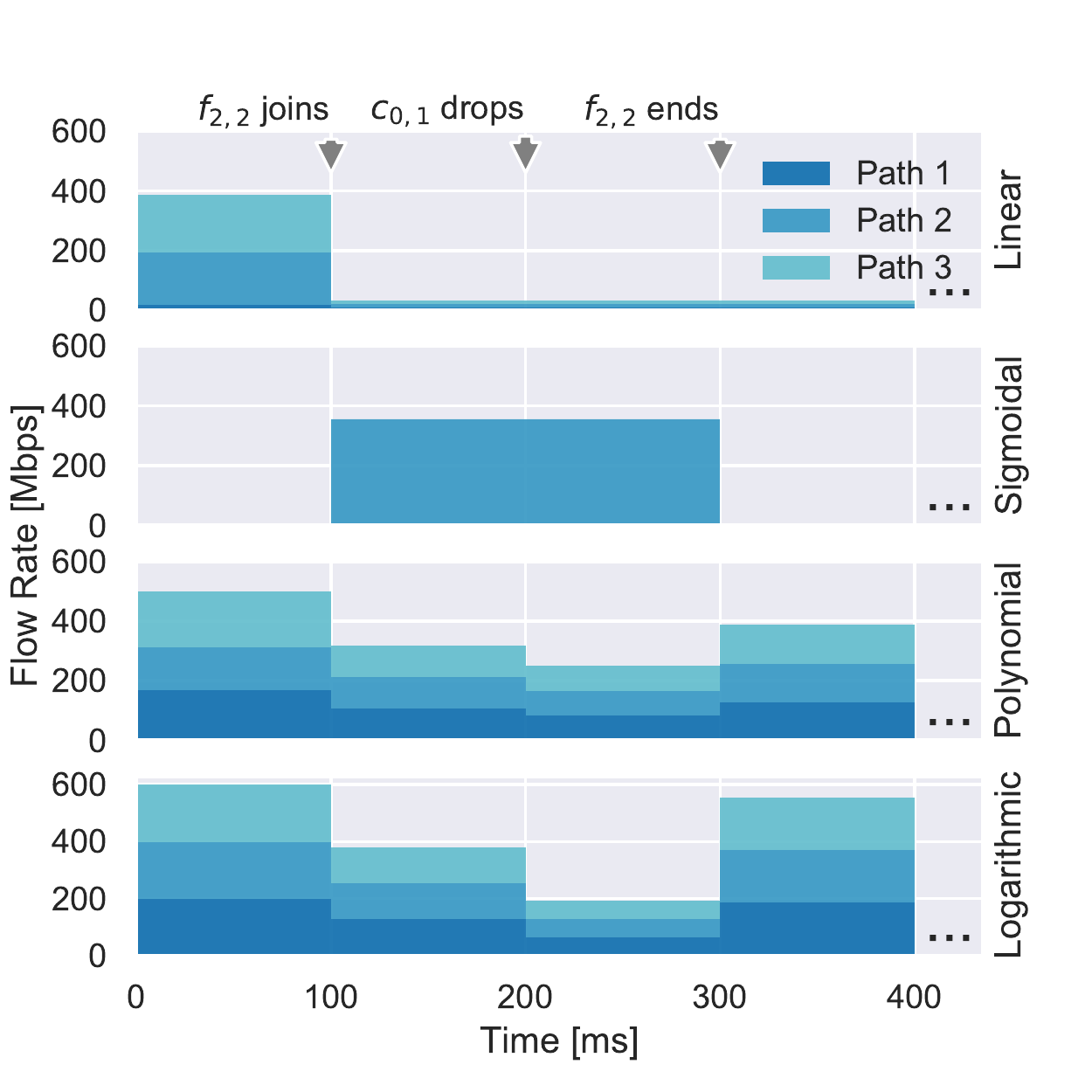}
 	\caption{Rate allocations performed by {\scshape Delmu} for flows of different slices and paths over time in Topology 3 (see Fig.~\ref{fig:topos}), as a sequence of demand and capacity changes occur as labeled at the top of the figure. Numerical results.} 
 	\label{fig:rateTime}
\end{figure}

We complete this analysis by investigating the ability of the proposed {\sc Delmu} solution to handle network dynamics in sliced mm-wave backhaul settings, including changes in traffic demand due to e.g. on/off behaviour of user applications and variations in capacity triggered e.g. by occasional blockage on the mm-wave links. We consider Topology 3 in Fig.~\ref{fig:topos}, transporting a mix of flows with linear, polynomial, and logarithmic utility and different lifetimes, considering a 10~Mbps minimum level of service, in all cases when a flow is active. Precisely, in Fig.~\ref{fig:rateTime} we examine the time evolution of the throughput {\sc Delmu} allocates to flows on each slice, according to a sequence of events. In particular, flows subject to sigmoid utility start with 0~Mbps demands, whilst all flows of the other types on all path have each an initial demand of 200~Mbps. After 100 ms, a flow with sigmoid utility on path 2 (i.e. $f_{2,2}$) becomes active, adding a 400~Mbps demand to the network. At time 200 ms, partial link blockage occurs on the link between BS 0 and BS 1, causing the corresponding capacity $c_{0,1}$ to drop from 2,772 Mbps to 693 Mbps. $f_{2,2}$ finishes 100 ms later.

Observe in the figure that the {\sc Delmu} performs a correct allocation as soon as a change occurs and, given the millisecond scale inference times, the transition is almost instantaneous even at the 100 ms granularity. For instance, when $f_{2,2}$ joins, the allocation of network resources is immediately rearranged, so that the request of $f_{2,2}$ is mostly satisfied, whereas the rest of flows receive reduced rates. In this case, all the flows with linear utility are reduced to close to the minimum level of service, i.e. each to 11~Mbps rate. The drop in $c_{0,1}$ capacity at 200 ms leads to a significant degradation of the rates assigned to flows with polynomial and logarithmic utility, while the linear and sigmoid flows remain unaffected. Eventually, at 300 ms, when flow $f_{2,2}$ finishes, the rate of the flows with polynomial and logarithmic utility are increased, yet remain below the values assigned initially, due to the inferior $c_{0,1}$ capacity. Hence, the proposed {\sc Delmu} is suitable for highly dynamic backhaul environments, as it makes close to optimal inferences fast and is able to adapt to sudden changes.
\vspace{1mm}

\section{Related Work}
\label{sec:relatedwork}
\vspace{1mm}
In this section, we review previous work most closely related to our contribution, which touches upon network slicing, mm-wave backhauling, utility optimisation, and deep learning in networking.

\textbf{Network slicing.} 
Major 5G standardisation efforts put emphasis on the evolution towards sliced network architectures~\cite{3gpp23501,3gpp23799}
, and recent research highlights the benefits of sharing mobile infrastructure among virtual operators~\cite{samdanis:2016,sciancalepore2017mobile,sciancalepore2018onets}. In~\cite{samdanis:2016}, a slice broker concept that enables MIPs to manage dynamically the shared network resources is proposed. 
Based on this concept, a machine learning approach that addresses admission control in sliced networks is given in~\cite{sciancalepore2017mobile}. 
An online slice brokering solution is studied in~\cite{sciancalepore2018onets} with the goal of maximising the multiplexing gain in shared infrastructure. 
However, existing efforts do not address the diverse service requirements of different application scenarios.

\textbf{Mm-wave backhauling.} 
Mm-wave technology is recognised as a key enabler of multi-Gbps connectivity. Dehos~\emph{et al.} study the feasibility of employing mm-wave bands in access and backhaul networks, and highlight the significant throughput gain achievable at mm-wave frequencies as compared with microwave bands~\cite{dehos2014millimeter}. 
Hur \emph{et al.} propose a beam alignment scheme specifically targeting mm-wave backhauling scenarios and study the wind effect on the performance of backhaul links~\cite{hur2013millimeter}. 
Sim~\emph{et al.} propose a decentralised learning based medium access protocol for multi-hop mm-wave networks~\cite{sim2016learning}. In~\cite{li2016wihaul}, the authors advocate a max-min fair flow rate and airtime allocation scheme for mm-wave backhaul networks. These efforts however do not consider network utility and disregard sliced multi-service settings.

\textbf{Network utility maximisation (NUM).} 
With growing popularity of inelastic traffic, optimising a mix of both concave and non-concave utilities has been studied~\cite{Fazel2005,Hande2007,Chen2011}. 
Fazel~\emph{et al.} propose a sum-of-square method to solve non-concave NUM problems that tackle primarily polynomial utility~\cite{Fazel2005}.
Hande~\emph{et al.} study the sufficient conditions for the standard price-based (sub-gradient based dual) approach to converge to global optima with zero duality gap, which relies on capacity provisioning~\cite{Hande2007}. 
Chen~\emph{et al.} consider NUM with mixed elastics and inelastic traffic, and develop a heuristic method to approximate the optimal~\cite{Chen2011}. Recent work investigates convex relaxation of polynomial NUM and employs distributed heuristics to approximate the global optimal~\cite{Wang2017}, and Udell and Boyd define a general class of non-convex problems as sigmoidal programming and propose an approximation algorithm~\cite{Udell2013}.
The limitation of these heuristics lies within their convergence times that is in the order of seconds, which can hardly meet the latency requirements of 5G networks. In contrast, our deep learning approach infers close to optimal rate allocations within milliseconds.

\textbf{Deep learning in networking.} 
With the increase in computational power and data sets availability, a range of deep learning applications in the computer and communications networking domain are emerging~\cite{zhang2018deep}. A fully-connected neural network is used in~\cite{kato-wcm} to find optimal routes in wired/wireless heterogeneous
networks. Zhang \emph{et al.} employ a dedicated CNNs to infer fine-grained mobile traffic consumption from coarse traffic aggregates~\cite{Zhang2017ZipNet}, improving measurement resolution by up to 100$\times$ while maintaining hight accuracy. CNNs have also employed been in \cite{zhang2018long}, where the authors incorporate a 3D-CNN structure into a spatio-temporal neural network, to perform long-term mobile traffic forecasting. To the best of our knowledge, our work is the first that uses deep learning to solve utility optimisation problems in sliced backhauls.


\section{Conclusions}
\label{sec:conclusions}
In this paper we tackled utility optimisation in sliced mm-wave networks by proposing {\sc Delmu}, a deep learning approach that learns correlations between traffic demands and optimal rate allocations. We specifically deal with scenarios where traffic is subject to conflicting requirements and maximise non-concave utility functions that reconcile all services, while overcoming the inherent complexity of the problems posed. We demonstrated that the proposed convolutional neural network attains up to 62\% utility gains over a greedy approach, infers close to optimal allocation solutions within orders of magnitude shorter runtimes as compared to global search, and responds quickly to network dynamics.

\bibliographystyle{unsrt}
\bibliography{references}

\end{document}